\documentclass[prd, twocolumn, showpacs, superscriptaddress, amsmath, amssymb]{revtex4}

\usepackage{graphicx}
\usepackage{bm}

\begin{document}

\title{Neutron to Mirror-Neutron Oscillations in the Presence of Mirror Magnetic Fields}

\def\ECTUM{Excellence Cluster `Universe', \TUM}
\def\GUM{Johannes--Gutenberg--Universit\"at, D--55128 Mainz, Germany}
\def\HNINP{Henryk Niedwodnicza\'nski Institute for Nuclear Physics, 31--342 Cracow, Poland}
\def\ILL{Institut Laue--Langevin, F--38042 Grenoble Cedex, France}
\def\JENA{Department of Neurology, Friedrich--Schiller--University, D--07740 Jena, Germany}
\def\JINR{JINR, 141980 Dubna, Moscow region, Russia}
\def\JUC{Marian Smoluchowski Institute of Physics, Jagiellonian University, 30--059 Cracow, Poland}
\def\KCGUM{Institut f\"ur Kernchemie, \GUM}
\def\KULEUVEN{Katholieke Universiteit Leuven, Instituut voor Kern-- en Stralingsfysica, Celestijnenlaan 200D, B--3001 Leuven, Belgium}
\def\LPC{LPC Caen, ENSICAEN, Universit\'e de Caen, CNRS/IN2P3, F--14050 Caen, France}
\def\LPSC{LPSC, Universit\'e Joseph Fourier Grenoble 1, CNRS/IN2P3, Institut National Polytechnique de Grenoble 53, F--38026 Grenoble Cedex, France}
\def\PGUM{Institut f\"ur Physik, \GUM}
\def\PSI{Paul Scherrer Institut (PSI), CH--5232 Villigen PSI, Switzerland}
\def\RAL{Rutherford Appleton Laboratory, Chilton, Didcot, Oxon OX11 0QX, United Kingdom}
\def\SUSSEX{Department of Physics and Astronomy, University of Sussex, Falmer, Brighton BN1 9QH, United Kingdom}
\def\TUM{Technische Universit\"at M\"unchen, D--85748 Garching, Germany}
\def\UNIFR{University of Fribourg, CH--1700, Fribourg, Switzerland}

\author{I.~Altarev}      \affiliation{\TUM}
\author{C.~A.~Baker}      \affiliation{\RAL}
\author{G.~Ban}          \affiliation{\LPC}
\author{K.~Bodek}        \affiliation{\JUC}
\author{M.~Daum}         \altaffiliation{Also at TU M\"unchen and University of Virginia} \affiliation{\PSI}
\author{P.~Fierlinger}   \affiliation{\ECTUM}
\author{P.~Geltenbort}   \affiliation{\ILL}
\author{K.~Green}   \affiliation{\RAL} \affiliation{\SUSSEX}
\author{M.~G.~D.~van~der~Grinten}   \affiliation{\RAL} \affiliation{\SUSSEX}
\author{E.~Gutsmiedl}    \affiliation{\TUM}
\author{P.~G.~Harris}       \affiliation{\SUSSEX}
\author{R.~Henneck}      \affiliation{\PSI}
\author{M.~Horras}       \affiliation{\ECTUM} \affiliation{\PSI}
\author{P.~Iaydjiev}       \altaffiliation{On leave from INRNE, Sofia, Bulgaria} \affiliation{\RAL}
\author{S.~Ivanov}       \altaffiliation{On leave from PNPI, St Petersburg, Russia} \affiliation{\RAL}
\author{N.~Khomutov}     \affiliation{\JINR}
\author{K.~Kirch}        \email[]{klaus.kirch@psi.ch} \affiliation{\PSI}
\author{S.~Kistryn}      \affiliation{\JUC}
\author{A.~Knecht}       \email[]{a.knecht@psi.ch} \altaffiliation{also at University of Z\"urich} \affiliation{\PSI}
\author{P.~Knowles}      \affiliation{\UNIFR}
\author{A.~Kozela}       \affiliation{\HNINP}
\author{F.~Kuchler}      \affiliation{\ECTUM}
\author{M.~Ku\'zniak}    \altaffiliation{Now at Queen's University, Kingston ON, Canada} \affiliation{\JUC} \affiliation{\PSI}
\author{T.~Lauer}        \affiliation{\KCGUM}
\author{B.~Lauss}        \affiliation{\PSI}
\author{T.~Lefort}       \affiliation{\LPC}
\author{A.~Mtchedlishvili}  \affiliation{\PSI}
\author{O.~Naviliat-Cuncic} \affiliation{\LPC}
\author{S.~Paul}         \affiliation{\TUM}
\author{A.~Pazgalev}     \affiliation{\UNIFR}
\author{J.~M.~Pendlebury}     \affiliation{\SUSSEX}
\author{G.~Petzoldt}     \affiliation{\PSI}
\author{E.~Pierre}       \affiliation{\LPC} \affiliation{\PSI}
\author{C.~Plonka-Spehr} \affiliation{\KCGUM}
\author{G.~Qu\'em\'ener} \affiliation{\LPSC}
\author{D.~Rebreyend}    \affiliation{\LPSC}
\author{S.~Roccia}       \affiliation{\LPSC}
\author{G.~Rogel}        \affiliation{\ILL} \affiliation{\LPC}
\author{N.~Severijns}    \affiliation{\KULEUVEN}
\author{D.~Shiers}    \affiliation{\SUSSEX}
\author{Yu.~Sobolev}     \affiliation{\PGUM}
\author{R.~Stoepler}     \affiliation{\TUM}
\author{A.~Weis}         \affiliation{\UNIFR}
\author{J.~Zejma}        \affiliation{\JUC}
\author{J.~Zenner}       \affiliation{\KCGUM}
\author{G.~Zsigmond}     \affiliation{\PSI}

\date{\today}

\begin{abstract} We performed ultracold neutron (UCN) storage measurements to search for additional losses due to neutron ($n$) to mirror-neutron ($n'$) oscillations as a function of an applied magnetic field $B$. In the presence of a mirror magnetic field $B'$, UCN losses would be maximal for $B\approx B'$. We did not observe any indication for $nn'$ oscillations and placed a lower limit on the oscillation time of $\tau_{nn'} > 12.0\,\text{s}$ at 95\% C.L. for any $B'$ between 0 and 12.5~$\mu\text{T}$.
\end{abstract}


\maketitle

\section{Introduction}
The idea of restoring global parity symmetry by introducing mirror particles dates back to Lee and Yang~\cite{lee56}. In~\cite{kob66}, this idea has been significantly expanded and was later adapted to the framework of the Standard Model of particle physics~\cite{foo91}. A recent review can be found in~\cite{oku06}. Interactions between ordinary and mirror particles are possible, e.g., they both feel gravity, making mirror matter a viable candidate for dark matter~\cite{bli82, foo04, ber05,foo08}. Besides gravity, new interactions could lead to mixings between neutral particles and their mirror partners.

Fast $nn'$ oscillations were introduced in~\cite{ber06} to explain the existence of ultra-high energy cosmic rays, based on a crude limit on the oscillation time $\tau_{nn'}\gtrsim 1\,\text{s}$. This weak limit was one of the motivations to perform a first dedicated measurement, which resulted in a lower limit of $\tau_{nn'} > 103$ s (95\% C.L.)~\cite{ban07}. The experiment relied on comparing the numbers of stored UCN remaining after a certain storage time for zero magnetic field and for an applied magnetic field $B$ of several $\mu$T~\cite{pok06}. Only for $B\approx 0$ would the ordinary and mirror state be degenerate and $nn'$ oscillations could occur leading to an additional loss of stored UCN. Shortly thereafter, an improved result of $\tau_{nn'} > 414$~s (90\% C.L.) was reported~\cite{ser08} and further improved to $\tau_{nn'} > 448$~s (90\% C.L.)~\cite{ser08b}.

So far, the limits were obtained assuming a negligible mirror magnetic field $B'$, except from an attempt in \cite{ser08b} for mirror magnetic fields in the range 0 to 1.2~$\mu$T. Here, we report the first systematic search for $nn'$ oscillations allowing for the presence of $B'$. The basic measurement principle remains unchanged with the exception of scanning $B$ in order to find a resonance of maximal UCN losses at $B\approx B'$ instead of $B\approx 0$. The limits on $B'$ from, e.g., a limit on the amount of mirror matter inside the earth~\cite{ign00} are very weak. Photon--mirror-photon mixings could possibly provide an efficient mechanism to capture mirror matter in the earth allowing for $B'$ of several $\mu$T~\cite{ber08}. Mirror magnetic fields not bound to the earth are also conceivable and would additionally lead to daily modulations in the UCN counts -- an unmistakable signature of a possible origin of $B'$. In the following, we will first introduce the theory of $nn'$ oscillations in the presence of $B'$, describe the measurements and conclude with the two analyses conducted: i) the search for daily modulations and ii) the search for a resonance.

\section{$nn'$ Oscillations in the Presence of a Mirror Magnetic Field}
For the calculation of the $nn'$ oscillation probability with finite $B'$, we follow the arguments of~\cite{ber08}. Defining $2\hbar\bm{\omega} \equiv \mu_n\bm{B}$ and $2\hbar\bm{\omega'} \equiv \mu_n\bm{B'}$ and introducing the oscillation time $\tau_{nn'}$ and the Pauli matrices $\bm{\sigma}$, the transition from the ordinary to the mirror state (and vice versa) is described by the interaction hamiltonian
\begin{equation}
 \mathcal{H} = \hbar \left( \begin{array}{cc} 2 \bm{\omega}\cdot \bm{\sigma} & \tau_{nn'}^{-1} \\ \tau_{nn'}^{-1} & 2 \bm{\omega'}\cdot \bm{\sigma} \end{array} \right) \,.
\end{equation}
Defining a coordinate system with $\bm{b} = (0,0,b)$, $b=|\bm{\omega}+\bm{\omega'}|$, and $\bm{a} = (a_x, 0,a_z)$, $a_x = 2|\bm{\omega}\times\bm{\omega'}| / |\bm{\omega}+\bm{\omega'}|$, $a_z = (\bm{\omega}^2 - \bm{\omega'}^2) / |\bm{\omega}+\bm{\omega'}|$, leads to the $4\times 4$ matrix
\begin{equation}
 \mathcal{H} = \hbar \left(\begin{array}{cccc} b-a_z & -a_x & \tau_{nn'}^{-1} & 0 \\
 -a_x & -b+a_z & 0 & \tau_{nn'}^{-1} \\
 \tau_{nn'}^{-1} & 0 & b+a_z & a_x \\
 0 & \tau_{nn'}^{-1} & a_x & -b-a_z \end{array} \right)\,.
\end{equation}
$\mathcal{H}$ can be diagonalised using a transformation matrix with mixing angles fulfilling $\tan{2\theta} = 1 / (a_z\tau_{nn'})$, $\tan{2\phi} = a_x / (b-\tilde{a}_z)$, and $\tan{2\phi'} = a_x / (b+\tilde{a}_z)$ with $\tilde{a}_z = a_z\sqrt{1+1/(a_z\tau_{nn'})^2}$~\cite{ber08}. The eigenvalues of $\mathcal{H}$ are $\pm 2\tilde{\omega}$ and $\pm 2\tilde{\omega}'$ given by $2\tilde{\omega} = a_x \sin{2\phi}+(b-\tilde{a}_z)\cos{2\phi}$ and $2\tilde{\omega}' = a_x \sin{2\phi'}+(b+\tilde{a}_z)\cos{2\phi'}$. The time dependent probability for the transition from $n$ to $n'$ is then given by
\begin{eqnarray}\label{Eq:nn'_prob}
 P_{nn'}(t) & = & \sin^2(2\theta) \left[ \cos^2(\phi - \phi') \sin^2\left(t / \tau_- \right) \right. \nonumber\\ & & \left. + \sin^2(\phi - \phi') \sin^2\left(t / \tau_+ \right) \right] \,,
\end{eqnarray}
where $\tau_{\pm} = |\tilde{\omega}\pm\tilde{\omega}'|^{-1}$ are the effective oscillation times. The oscillation probability depends on the magnitude of $B$ and $B'$, the direction of $B'$ given by the angle $\beta$ relative to the up-direction of $B$ (see below), the oscillation time $\tau_{nn'}$, and the time $t$.

During the storage of UCN inside a chamber, the relevant time $t$ is the free flight time $t_f$ between wall collisions in which the wave function is projected onto its pure $n$ or $n'$ state. The loss rate of UCN due to $nn'$ oscillations is thus given as
\begin{equation}
 R_{t_s} = f_c P_{nn'} = \frac{1}{\langle t_f \rangle_{t_s}} \langle P_{nn'}(t_f) \rangle_{t_s} \,,
\end{equation}
where $f_c$ denotes the collision frequency and $\langle \ldots \rangle_{t_s}$ the averaging over the distribution of free flight times $t_f$ during the storage time $t_s$.

There are two distinct regions for the evaluation of the $nn'$ oscillation probability. The first is the off--resonance region. From evaluations of Eq.~(\ref{Eq:nn'_prob}), this holds for $|B-B'|>0.4\,\mu\text{T}$. In this region, the time dependent terms in Eq.~(\ref{Eq:nn'_prob}) oscillate quickly and average to $1/2$ over the $t_f$ distribution. The loss rate is then expressed explicitly as
\begin{equation}\label{Eq:nn'_prob_off_res}
 R_{t_s}^{\text{off}} = \frac{1}{\langle t_f \rangle_{t_s}} \frac{B'^2+B^2+2B'B\cos{\beta}}{(B'^2-B^2)^2} \frac{2\hbar^2}{\mu_n^2\tau_{nn'}^2}\,.
\end{equation}
On--resonance, $|B-B'|<0.4\,\mu\text{T}$, the first term in Eq.~(\ref{Eq:nn'_prob}) dominates for most of the parameter space. For that part of the parameter space, we have $\phi \approx \phi'$ and, since $t / \tau_-$ is small, $\sin^2\left(t / \tau_- \right) \approx (t / \tau_-)^2$. Therefore, we can replace $t$ in Eq.~(\ref{Eq:nn'_prob}) by $\sqrt{\langle t_f^2 \rangle_{t_s}}$ and write the loss rate as
\begin{equation}\label{Eq:nn'_prob_on_res}
 R_{t_s}^{\text{on}} \approx \frac{1}{\langle t_f \rangle_{t_s}} P_{nn'}(\sqrt{\langle t_f^2 \rangle_{t_s}})
\end{equation}
The validity of Eq.~(\ref{Eq:nn'_prob_on_res}) was checked by comparing to a full averaging over a realistic $t_f$ distribution. Deviations were less than 1\%. Anyhow, our final limit is based on calculations using Eq.~(\ref{Eq:nn'_prob_off_res}).

In order to obtain the values for $\langle t_f \rangle_{t_s}$ and $\sqrt{\langle t_f^2 \rangle_{t_s}}$, a detailed Monte Carlo simulation of the experiment was performed using GEANT4UCN~\cite{atc05} with parameters tuned to reproduce experimental data (such as characteristic time constants for filling, emptying, or storage). The $t_f$ distributions were obtained from the time of the reflections of individual trajectories inside the storage chamber. Results are given in Table~\ref{Tab:tf_sim} for the two storage times $t_s$ used in the measurements. We varied the parameters of the simulation in ranges still reproducing the experimental data to assess the systematic uncertainties.

The number of surviving UCN after storage is
\begin{equation}\label{Eq:UCN_cts}
 N(t_s^*) = N'_{0,t_s} \exp{\left(- R_{t_s} t_s^*\right)} \,
\end{equation}
where $N'_{0,t_s}$ is the initial number of UCN reduced by the usual losses during storage, and $t^*_s$ is the effective storage time for the UCN, including not only the time when the neutrons are fully confined, $t_s$, but also the effects of storage chamber filling and emptying. The values for $t_s^*$ are given in Table~\ref{Tab:tf_sim}.

In the case of a mirror magnetic field not bound to the earth, the observed neutron counts could be modulated with a period corresponding to a sidereal day ($d_{sid}$ = 23.934~h) as the angle $\beta$ would be modulated. For the off--resonance case, the observed counts are then given by $N(t) = \mathcal{C} + \mathcal{A}\frac{t_s^*}{\langle t_f \rangle_{t_s}}\cos\left(2\pi (t-t_0) / d_{sid}\right)$ with
\begin{eqnarray} \label{Eq:cts_modulation}
 \mathcal{C} & \approx & N_0'\left( 1 - \frac{t_s^*}{\langle t_f \rangle_{t_s}} \frac{B'^2+B^2}{(B'^2-B^2)^2} \frac{2\hbar^2}{\mu_n^2\tau_{nn'}^2} \right. \nonumber \\
 & & \left.\mp \frac{t_s^*}{\langle t_f \rangle_{t_s}} \frac{B B'_\shortparallel}{(B'^2-B^2)^2} \frac{4\hbar^2}{\mu_n^2\tau_{nn'}^2} \sin\lambda\right) \,, \nonumber \\
 \mathcal{A} & \approx & \mp N_0' \frac{B B'_\bot}{(B'^2-B^2)^2} \frac{4\hbar^2}{\mu_n^2\tau_{nn'}^2} \cos\lambda \,.
\end{eqnarray}
$B'_\shortparallel$ and $B'_\bot$ are the components of $B'$ parallel and perpendicular to the earth's rotation axis, $\lambda$ the latitude at the experimental site, $t_0$ the phase, and the $-(+)$ sign stands for magnetic field up (down).

\begin{table}
 \begin{center}
 \caption{\label{Tab:tf_sim} Results for $\langle t_f \rangle_{t_s}$ and $\sqrt{\langle t_f^2 \rangle_{t_s}}$ using Monte Carlo calculations and the effective storage times $t_s^*$. The values at the right side of the arrow denote the values used in the calculations in order to obtain a conservative result.}
 \begin{tabular}{c c c}
 \hline \hline
 $t_s\,[\,{\rm s}\,]$ & 75 & 150\\
 \hline
 $\langle t_f \rangle_{t_s}\,[\,{\rm s}\,]$ & 0.0403(4) $\rightarrow$ 0.0407 & 0.0442(4) $\rightarrow$ 0.0446\\
 $\sqrt{\langle t_f^2\rangle_{t_s}}\,[\,{\rm s}\,]$ & 0.0532(5) $\rightarrow$ 0.0527  & 0.0586(6) $\rightarrow$ 0.0580 \\
 $t_s^*\,[\,{\rm s}\,]$ & 98(3) $\rightarrow$ 95 & 173(3) $\rightarrow$ 170\\
 \hline\hline
 \end{tabular}
 \end{center}
 \vspace{-0.3cm}
\end{table}

\section{Measurements}
The UCN storage experiments were conducted at the PF2-EDM beamline~\cite{ste86} at the Institut Laue-Langevin (ILL) using the apparatus for the search of the neutron electric dipole moment~\cite{bak06b}. The main features of the apparatus are: i) the possibility to efficiently store UCN in vacuum in a chamber made from deuterated polystyrene~\cite{bod08} and diamond-like carbon and ii) the surrounding 4-layer Mu-metal shield together with an internal magnetic field coil that allowed to set and maintain magnetic fields with a precision of $\sim 0.1\,\mu\text{T}$. A typical measurement cycle consisted of filling unpolarised UCN for 40~s into the storage chamber of 21 litres, confining the UCN for 75~s (150~s) and subsequently counting $\sim 38000$ ($\sim 24000$) UCN over 40~s in a $^3$He detector~\cite{footnote_3He_det}. For a given magnetic field value, we always performed 8 cycles with a storage time of 75~s and 8 cycles with a storage time of 150~s. After these 16 cycles, the magnetic field direction was changed from up to down and measured again for 16 cycles. The averages of the different $B$ field settings, applied randomly, were 0, $2.5\,\mu\text{T}$, $5\,\mu\text{T}$, $7.5\,\mu\text{T}$, $10\,\mu\text{T}$, $12.5\,\mu\text{T}$. Before doing a zero field measurement, the 4-layer magnetic shield was demagnetised resulting in $B<50\,\text{nT}$. In total, data taken continuously over approximately 110 hours was used for the analysis.

\section{Normalisation of the UCN Data}
The data showed a trend to higher UCN counts over the course of the measurement period. The increase amounted to $\sim$2.5\% for 75 s storage time and $\sim$5\% for 150 s storage. We attribute this increase to slowly improving vacuum conditions inside the chamber. A combined fit to both data sets was performed with the function
\begin{equation}\label{Eq:outgassing_fit}
 f_{t_s}(t) = N_{t_s} \exp{\left(-C_p t_s e^{-\frac{t}{\tau_p}} - C_R t_s^2 e^{-\frac{t}{\tau_R}}\right)}
\end{equation}
with two normalisation constants $N_{75}$ and $N_{150}$ and two constants proportional to a decreasing overall pressure $C_p$ (with a characteristic time $\tau_p$) and a decreasing outgassing rate $C_R$ (characteristic time $\tau_R$) of the storage chamber, which is sealed off from the pumps during storage. The $\chi^2$ per degree of freedom, 1386/1204, is satisfactory. Assuming a UCN loss cross section per molecule of $\mathcal{O}$(10 b), the fitted constants $C_p$ and $C_R$ translate into an initial pressure of $\mathcal{O}(10^{-3}$ mbar) and an initial outgassing rate of $\mathcal{O}(10^{-7} \text{mbar}\,\text{l}\,\text{s}^{-1}\,\text{cm}^{-2})$ which both seem realistic \cite{bod08}. We normalised the UCN counts for a given cycle by the prediction of Eq.~(\ref{Eq:outgassing_fit}) and slightly increased the statistical error by adding the fit error in quadrature. Residual drifts ($\lesssim 0.5\%$ over several hours) showed a weak correlation to the ILL reactor power. Their effect on the final result is negligible.

\section{Analysis}
We conducted two different types of analyses: i) The search for a modulation in the UCN counts and ii) the search for a resonance in the UCN counts as a function of $B$. It is clear from Eqs.~(\ref{Eq:cts_modulation}) and (\ref{Eq:nn'_prob_off_res}) that the resonance analysis will always be sensitive to $nn'$ oscillations regardless of the origin of the mirror magnetic field and possible modulation periods whereas the modulation analysis is not. In Eq.~(\ref{Eq:nn'_prob_off_res}), $\cos{\beta}$ will either be a fixed value or the average over a modulated $\cos{\beta}$. Additionally, the amplitude of the modulation tends to zero for small $B'$ and the constant term $\mathcal{C}$ of the oscillation probability is for all parameters larger or equal to the modulated part $\mathcal{A}$ ($B'^2+B^2 \ge 2B'B\cos{\beta}$). Given the same statistics and no systematic errors from averaging over longer periods, the resonance analysis will always yield tighter constraints on $\tau_{nn'}$ than the modulation analysis. As a means of crosschecking and discovering the possible origin of $B'$, both types of analyses have been performed.

\subsection{Search for a Daily Modulation}
\begin{table}
 \begin{center}
 \caption{\label{Tab:mod_ampl} Results of the fits using Eq.~(\ref{Eq:asym_modulation}) to the up/down asymmetries $A$ for the five different magnetic field values and the upper limits on the amplitude of a daily modulation $\mathcal{A}_\text{lim}$ at 95\% C.L. and for any value of the phase $t_0$.}
 \begin{tabular}{c c c c c}
 \hline \hline
 $B$ [$\mu$T] & $\mathcal{A}$ $\times 10^{7}$& $t_0$ [h] &
 $\chi^2$/dof &
 $\mathcal{A}_\text{lim}$ $\times 10^{7}$ \\
 \hline
 2.5 & $1.3\pm 1.8$ & $11.7\pm 9.4$ & 6.53/10 & 6.6 \\
 5 & $2.4\pm 2.3$ & $14.6\pm 3.0$ & 5.92/10 & 6.4 \\
 7.5 & $3.5\pm 2.4$ & $0.3\pm 2.0$ & 5.52/10 & 7.6 \\
 10 & $0.6\pm 1.9$ & $11.6\pm 12.6$ & 18.05/12 & 5.0 \\
 12.5 & $1.0\pm 1.7$ & $17.1\pm 9.8$ & 10.13/12 & 5.0 \\
 \hline \hline
 \end{tabular}
 \end{center}
 \vspace{-0.3cm}
\end{table}

In order to search for a modulation without being affected by the slow residual drifts present in the normalised UCN data, we calculated the up/down-asymmetries in the UCN counts $A = (N_\uparrow - N_\downarrow) / (N_\uparrow + N_\downarrow)$ from the two subsequent (within $\sim$1 h) measurements at $B$ field up and down. The two asymmetry data sets for 75~s and 150~s were separately normalised in order to have zero weighted means. A modulation in the UCN counts would show up in the asymmetry with the same amplitude $\mathcal{A}$ as given in Eq.~(\ref{Eq:cts_modulation}):
\begin{equation}\label{Eq:asym_modulation}
 A(t) = \mathcal{A}\frac{t_s^*}{\langle t_f \rangle_{t_s}}\cos\left(\frac{2\pi}{d_{sid}}(t-t_0)\right) \,.
\end{equation}
We searched for a modulation in the 5 data sets of different $B$ ($2.5\,\mu\text{T}$, $5\,\mu\text{T}$, $7.5\,\mu\text{T}$, $10\,\mu\text{T}$, and $12.5\,\mu\text{T}$) by fitting Eq.~(\ref{Eq:asym_modulation}) to the data. None of the fits showed a significant modulation. Limits on the amplitude were calculated performing a frequentist confidence level analysis along the lines of \cite{alt09}. The results of the fits and the corresponding limits are listed in Table~\ref{Tab:mod_ampl}.

\subsection{Search for a Resonance}

\begin{figure}
 \includegraphics[width=\linewidth, angle=0]{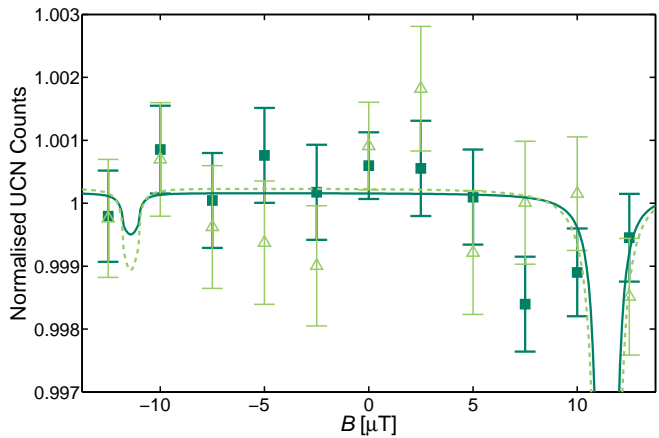}
 \caption{\label{Fig:resonance} (colour online) Combined fit to the normalised UCN counts as a function of applied magnetic field $B$ for 75 s (dark green squares and solid line) and 150 s (light green triangles and dashed line). Positive (negative) $B$ values correspond to $B$ field up (down).}
 \vspace{-0.3cm}
\end{figure}

\begin{figure}
 \includegraphics[width=\linewidth, angle=0]{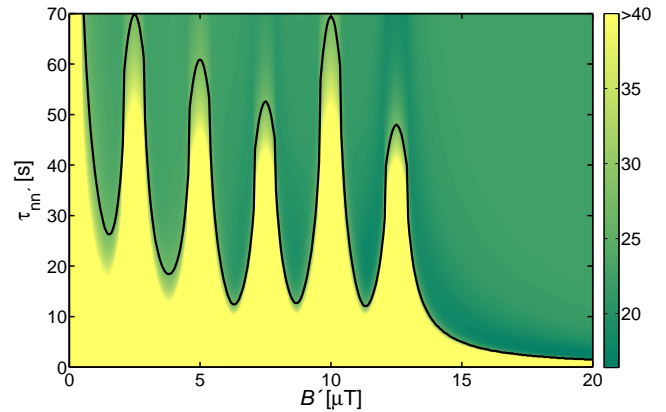}
 \caption{\label{Fig:tau_limit} (colour online) Contour plot of the minimal $\chi^2$ at the point $(B',\,\tau_{nn'})$. The solid line denotes the 95\% C.L. contour line for an exclusion of $\tau_{nn'}$. We evaluated a lower limit on $\tau_{nn'}$ at the minimum of this contour for $B'$ between 0 and 12.5~$\mu\text{T}$.}
 \vspace{-0.3cm}
\end{figure}

In order to search for a resonance in the loss rate at the point $B\approx B'$, we averaged all normalised UCN counts for individual $B$ field settings (thereby averaging out any remaining long term drifts) and plotted the results as a function of $B$ (see Fig.~\ref{Fig:resonance}). A combined fit to the two data sets was performed using Eq.~(\ref{Eq:UCN_cts}) with the following free parameters: two normalisation constants $N'_{75}$ and $N'_{150}$, the magnitude of $B'$, the angle $\beta$, and the oscillation time $\tau_{nn'}$. The value for $B'$ was constrained to lie in the region $0\ldots 12.5\,\mu\text{T}$ as only in that region we would have unambiguous evidence for a possible resonance. The relevant, fitted parameters are $B'=11.4\,\mu\text{T}$, $\beta=25.3\,\text{deg}$, and $\tau_{nn'}=21.9\,\text{s}$. The $\chi^2$ per degree of freedom ($\chi^2/\text{dof}=17.86/17$) is comparable to the one obtained by fitting a constant to the data ($\chi^2/\text{dof}=22.72/21$). There is therefore no evidence of a mirror magnetic field present at the site of the experiment and the data were used to set a limit on $\tau_{nn'}$ for mirror magnetic fields between 0 and 12.5 $\mu$T. To do so, the minimal $\chi^2$ at the points $(B',\,\tau_{nn'})$ was calculated by fitting the remaining free parameters $N'_{75}$, $N'_{150}$, and $\beta$ (see Fig.~\ref{Fig:tau_limit}). The 95\% C.L. contour corresponds to $\chi^2=27.59$, the 95\% C.L. for a $\chi^2$ distribution with 17 degrees of freedom. Figure~\ref{Fig:tau_limit} also shows the loss of sensitivity to $nn'$ oscillations for $B'$ fields outside the range of applied magnetic fields. We evaluated a lower limit on the oscillation time as the minimal $\tau_{nn'}$ on this contour for $B'$ between 0 and 12.5~$\mu$T:
\begin{equation}
 \tau_{nn'} > 12.0\,\text{s (95\% C.L.)}
\end{equation}
The 0.1 $\mu$T precision on individual non-zero $B$ field values leads in principle to a systematically improved limit. The improvement could not be quantified exactly, but it is estimated to be less than 1 s, and was not included in the result. Additionally, we improve our previous limit on $\tau_{nn'}$ for negligible $B'$ at the intercept of the exclusion contour line in Fig. \ref{Fig:tau_limit} with $B'=0$: $\tau_{nn'} > 141\,\text{s (95\% C.L.)}$.

\begin{acknowledgments}
We are grateful to the ILL staff for providing us with excellent running conditions and in particular acknowledge the outstanding support of T.~Brenner. We also benefitted from the technical support throughout the collaboration. The work is supported by grants from the Polish Ministry of Science and Higher Education, contract No. 336/P03/2005/28, and the Swiss National Science Foundation \#200020–111958.
\end{acknowledgments}

\vspace{-0.3cm}


\begin{thebibliography}{23}
\expandafter\ifx\csname natexlab\endcsname\relax\def\natexlab#1{#1}\fi
\expandafter\ifx\csname bibnamefont\endcsname\relax
  \def\bibnamefont#1{#1}\fi
\expandafter\ifx\csname bibfnamefont\endcsname\relax
  \def\bibfnamefont#1{#1}\fi
\expandafter\ifx\csname citenamefont\endcsname\relax
  \def\citenamefont#1{#1}\fi
\expandafter\ifx\csname url\endcsname\relax
  \def\url#1{\texttt{#1}}\fi
\expandafter\ifx\csname urlprefix\endcsname\relax\def\urlprefix{URL }\fi
\providecommand{\bibinfo}[2]{#2}
\providecommand{\eprint}[2][]{\url{#2}}

\bibitem[{\citenamefont{Lee and Yang}(1956)}]{lee56}
\bibinfo{author}{\bibfnamefont{T.~D.} \bibnamefont{Lee}} \bibnamefont{and}
  \bibinfo{author}{\bibfnamefont{C.~N.} \bibnamefont{Yang}},
  \bibinfo{journal}{Phys. Rev.} \textbf{\bibinfo{volume}{104}},
  \bibinfo{pages}{254} (\bibinfo{year}{1956}).

\bibitem[{\citenamefont{Kobzarev et~al.}(1966)\citenamefont{Kobzarev, Okun, and
  Pomeranchuk}}]{kob66}
\bibinfo{author}{\bibfnamefont{I.~Y.} \bibnamefont{Kobzarev}},
  \bibinfo{author}{\bibfnamefont{L.~B.} \bibnamefont{Okun}}, \bibnamefont{and}
  \bibinfo{author}{\bibfnamefont{I.~Y.} \bibnamefont{Pomeranchuk}},
  \bibinfo{journal}{Sov. J. Nucl. Phys} \textbf{\bibinfo{volume}{3}},
  \bibinfo{pages}{837} (\bibinfo{year}{1966}).

\bibitem[{\citenamefont{Foot et~al.}(1991)\citenamefont{Foot, Lew, and
  Volkas}}]{foo91}
\bibinfo{author}{\bibfnamefont{R.}~\bibnamefont{Foot}},
  \bibinfo{author}{\bibfnamefont{H.}~\bibnamefont{Lew}}, \bibnamefont{and}
  \bibinfo{author}{\bibfnamefont{R.~R.} \bibnamefont{Volkas}},
  \bibinfo{journal}{Phys. Lett. B} \textbf{\bibinfo{volume}{272}},
  \bibinfo{pages}{67} (\bibinfo{year}{1991}).

\bibitem[{\citenamefont{Okun}(2006)}]{oku06}
\bibinfo{author}{\bibfnamefont{L.~B.} \bibnamefont{Okun}},
  \bibinfo{journal}{arXiv:hep-ph/0606202v2},
  \bibinfo{journal}{Sov. Phys. Usp.} \textbf{\bibinfo{volume}{50}},
  \bibinfo{pages}{380} (\bibinfo{year}{2007}).

\bibitem[{\citenamefont{Blinnikov and Khlopov}(1982)}]{bli82}
\bibinfo{author}{\bibfnamefont{S.~I.} \bibnamefont{Blinnikov}}
  \bibnamefont{and} \bibinfo{author}{\bibfnamefont{M.}~\bibnamefont{Khlopov}},
  \bibinfo{journal}{Sov. J. Nucl. Phys} \textbf{\bibinfo{volume}{36}},
  \bibinfo{pages}{472} (\bibinfo{year}{1982}).

\bibitem[{\citenamefont{Foot}(2004)}]{foo04}
\bibinfo{author}{\bibfnamefont{R.}~\bibnamefont{Foot}}, \bibinfo{journal}{Int.
  J. Mod. Phys. D} \textbf{\bibinfo{volume}{13}}, \bibinfo{pages}{2161}
  (\bibinfo{year}{2004}).

\bibitem[{\citenamefont{Berezhiani et~al.}(2005)\citenamefont{Berezhiani et~al.}}]{ber05}
\bibinfo{author}{\bibfnamefont{Z.}~\bibnamefont{Berezhiani}},
  \bibnamefont{et~al.},
  \bibinfo{journal}{Int. J. Mod. Phys. D} \textbf{\bibinfo{volume}{14}},
  \bibinfo{pages}{107} (\bibinfo{year}{2005}).

\bibitem[{\citenamefont{Foot}(2008)}]{foo08}
\bibinfo{author}{\bibfnamefont{R.}~\bibnamefont{Foot}}, \bibinfo{journal}{Phys.
  Rev. D} \textbf{\bibinfo{volume}{78}}, \bibinfo{pages}{043529}
  (\bibinfo{year}{2008}).

\bibitem[{\citenamefont{Berezhiani and Bento}(2006)}]{ber06}
\bibinfo{author}{\bibfnamefont{Z.}~\bibnamefont{Berezhiani}} \bibnamefont{and}
  \bibinfo{author}{\bibfnamefont{L.}~\bibnamefont{Bento}},
  \bibinfo{journal}{Phys. Rev. Lett.} \textbf{\bibinfo{volume}{96}},
  \bibinfo{pages}{081801} (\bibinfo{year}{2006}).

\bibitem[{\citenamefont{Ban et~al.}(2007)\citenamefont{Ban et~al.}}]{ban07}
\bibinfo{author}{\bibfnamefont{G.}~\bibnamefont{Ban}},
  \bibnamefont{et~al.}, \bibinfo{journal}{Phys. Rev. Lett.}
  \textbf{\bibinfo{volume}{99}}, \bibinfo{pages}{161603}
  (\bibinfo{year}{2007}).

\bibitem[{\citenamefont{Pokotilovski}(2006)}]{pok06}
\bibinfo{author}{\bibfnamefont{Y.~N.} \bibnamefont{Pokotilovski}},
  \bibinfo{journal}{Phys. Lett. B} \textbf{\bibinfo{volume}{639}},
  \bibinfo{pages}{214} (\bibinfo{year}{2006}).

\bibitem[{\citenamefont{Serebrov
  et~al.}(2008{\natexlab{a}})\citenamefont{Serebrov
  et~al.}}]{ser08}
\bibinfo{author}{\bibfnamefont{A.~P.} \bibnamefont{Serebrov}},
  \bibnamefont{et~al.}, \bibinfo{journal}{Phys. Lett. B}
  \textbf{\bibinfo{volume}{663}}, \bibinfo{pages}{181}
  (\bibinfo{year}{2008}{\natexlab{a}}).

\bibitem[{\citenamefont{Serebrov
  et~al.}(2008{\natexlab{b}})\citenamefont{Serebrov
  et~al.}}]{ser08b}
\bibinfo{author}{\bibfnamefont{A.~P.} \bibnamefont{Serebrov}},
  \bibnamefont{et~al.}, \bibinfo{journal}{arXiv:0809.4902v2 [nucl-ex]}
  (\bibinfo{year}{2008}{\natexlab{b}}).

\bibitem[{\citenamefont{Ignatiev and Volkas}(2000)}]{ign00}
\bibinfo{author}{\bibfnamefont{A.~Y.} \bibnamefont{Ignatiev}} \bibnamefont{and}
  \bibinfo{author}{\bibfnamefont{R.~R.} \bibnamefont{Volkas}},
  \bibinfo{journal}{Phys. Rev. D} \textbf{\bibinfo{volume}{62}},
  \bibinfo{pages}{023508} (\bibinfo{year}{2000}).

\bibitem[{\citenamefont{Berezhiani}(2008)}]{ber08}
\bibinfo{author}{\bibfnamefont{Z.}~\bibnamefont{Berezhiani}},
  \bibinfo{journal}{arXiv:hep-ph/0804.2088v1}  (\bibinfo{year}{2008}).

\bibitem[{\citenamefont{Atchison et~al.}(2005)\citenamefont{Atchison et~al.}}]{atc05}
\bibinfo{author}{\bibfnamefont{F.}~\bibnamefont{Atchison}},
  \bibnamefont{et~al.},
  \bibinfo{journal}{Nucl. Instr. Meth. A} \textbf{\bibinfo{volume}{552}},
  \bibinfo{pages}{513} (\bibinfo{year}{2005}).

\bibitem[{\citenamefont{Steyerl et~al.}(1986)\citenamefont{Steyerl
  et~al.}}]{ste86}
\bibinfo{author}{\bibfnamefont{A.}~\bibnamefont{Steyerl}},
  \bibnamefont{et~al.}, \bibinfo{journal}{Phys. Lett. A}
  \textbf{\bibinfo{volume}{116}}, \bibinfo{pages}{347} (\bibinfo{year}{1986}).

\bibitem[{\citenamefont{Baker et~al.}(2006)\citenamefont{Baker
  et~al.}}]{bak06b}
\bibinfo{author}{\bibfnamefont{C.~A.} \bibnamefont{Baker}},
  \bibnamefont{et~al.}, \bibinfo{journal}{Phys. Rev. Lett.}
  \textbf{\bibinfo{volume}{97}}, \bibinfo{pages}{131801}
  (\bibinfo{year}{2006}).

\bibitem[{\citenamefont{Bodek et~al.}(2008)\citenamefont{Bodek et~al.}}]{bod08}
\bibinfo{author}{\bibfnamefont{K.}~\bibnamefont{Bodek}},
  \bibnamefont{et~al.}, \bibinfo{journal}{Nucl. Instr. Meth. A}
  \textbf{\bibinfo{volume}{597}}, \bibinfo{pages}{222} (\bibinfo{year}{2008}).

\bibitem[{foo({\natexlab{b}})}]{footnote_3He_det}
\bibinfo{note}{The UCN detector was manufactured by Strelkov et al. at the
  Joint Institute for Nuclear Research, Dubna, Russia.}

\bibitem[{\citenamefont{Altarev et~al.}(2009)\citenamefont{Altarev
  et~al.}}]{alt09}
\bibinfo{author}{\bibfnamefont{I.}~\bibnamefont{Altarev}},
  \bibnamefont{et~al.}, \bibinfo{journal}{arXiv:0905.3221v1 [nucl-ex]}  (\bibinfo{year}{2009}).

\end{thebibliography}

\end{document}